\documentclass[12pt,preprint]{aastex}

\shortauthors{Cui \& Smith}
\shorttitle{Cessation of X-ray Pulsation of GX 1+4}

\begin{document}

\title{Cessation of X-ray Pulsation of GX 1+4} 

\author{Wei Cui and Benjamin Smith} 
\affil{Department of Physics, Purdue University, West Lafayette, IN 47907} 
\affil{cui, smith110@physics.purdue.edu}

\begin{abstract}

We report results from our weekly monitoring campaign on the X-ray pulsar 
GX 1+4 with the {\em Rossi X-ray Timing Explorer} satellite. The spin-down 
trend of GX 1+4 was continuing, with the pulsar being at its longest 
period ever measured (about 138.7 s). At the late stage of the campaign, 
the source entered an extended faint state, when its X-ray (2-60 keV) flux 
decreased significantly to an average level of 
$\sim 3 \times 10^{-10}\mbox { }ergs\mbox{ }cm^{-2}\mbox{ }s^{-1}$. It was 
highly variable in the faint state; the flux dropped to as low as  
$\sim 3 \times 10^{-11} ergs\mbox{ }cm^{-2}\mbox{ }s^{-1}$. In several
observations during this period, the X-ray pulsation became undetectable. 
We can, therefore, conclude conservatively that the pulsed fraction, which 
is normally $\gtrsim$ 70\% (peak-to-peak), must have decreased drastically 
in those cases. This is very similar to what was observed of GX 1+4 in 1996 
when it became similarly faint in X-ray. In fact, the flux at which the 
cessation of X-ray pulsation first occurred is nearly the same as it was 
in 1996. We suggest that we have, once again, observed the propeller 
effect in GX 1+4, a phenomenon that is predicted by theoretical models 
of accreting X-ray pulsars.

\end{abstract}

\keywords{accretion, accretion disks --- stars: pulsars: individual 
(GX~1+4) --- X-rays: stars} 

\section{Introduction}

Accreting X-ray pulsars are neutron stars in a binary configuration. The
X-ray emission of such a system is powered by the accretion of matter 
from the companion star onto the neutron star. Therefore, unlike their 
isolated counterparts, neutron stars in a binary configuration are 
strongly influenced by the mass accretion process. 

In many cases, it is believed that an accretion disk is formed around the 
neutron star in an accreting pulsar. The disk is disrupted by the strong 
magnetic field of the neutron star at the magnetospheric radius, where 
plasmas in the disk are channeled to the polar regions of the star 
along the field lines, creating ``hot spots'' there. However, if the 
magnetospheric radius is larger than the co-rotation radius of the neutron 
star, the accreted matter hits the centrifugal barrier and is likely to be 
expelled from the system 
(Illarionov \& Sunyaev 1975; Wang \& Robertson 1985; Lovelace, Romanova, 
\& Bisnovatyi-Kogan 1999; Bogovalov 2001). This phenomenon is often 
referred to as the propeller effect. 

Observationally, the propeller effect may manifest itself directly in 
the cessation of pulsation of an accreting pulsar, when the mass 
accretion rate is sufficiently low (and thus the magnetosphere is 
sufficiently large). This was indeed first observed in two systems, 
GX 1+4 and GRO J1744-28 (Cui 1997), which possess contrasting magnetic 
properties. It was seen again recently, based on our monitoring 
observations of GX 1+4 during its transition to a rare X-ray faint 
state (which was very similar to the 1996 episode; Cui 1997). In this 
paper, we present the new data. We show that the results obtained are 
in good agreement with those of previous measurements and thus provide 
additional support for the earlier interpretation.

\section{Observations}

We monitored GX 1+4 weekly with the {\em Rossi X-ray Timing Explorer} (RXTE) 
over a two-year period (2001--2002). Typical observing times are 2--3 ks 
(roughly one {\em RXTE} orbit), although some longer observations were taken, 
in coordination with the {\em Chandra X-ray Observatory}, following the 
triggering of a Target-of-Opportunity program. For this work, we will focus 
on the {\em RXTE} observations taken in 2002, when a transition to a faint 
state occurred. The results of the {\em Chandra} observation will be 
discussed elsewhere.

There are two detectors aboard {\em RXTE} (Bradt, Rothschild, \& Swank 1993), 
the Proportional Counter Array (PCA) and the High-Energy X-ray Timing 
Experiment (HEXTE). The PCA consists of five nearly identical large area 
proportional counter units (PCUs). It has a total collecting area of about 
6500 $cm^2$ and covers a nominal energy range of 2--60 keV. The HEXTE has 
a total effective area of about 1600 $cm^2$ in two clusters. Each cluster
contains 4 detectors, although one of the detectors lost its spectral
capability very early in the mission. The HEXTE covers a wide energy range 
of about 15--250 keV, overlapping the PCA passband at low energies. Since 
the HEXTE clusters alternate between on-source and off-source fields, the 
effective exposure time (with dead-time subtracted) is always much less 
than the actual observing time.

Operational constraints require that some PCUs be turned off. Which 
PCUs are turned off varies from observation to observation. In our case, 
the number of PCUs that were operational varied from 2 to 5, with PCU 0 
and PCU 2 being on throughout the entire campaign. Because PCU 0 has
lost its front veto layer, which makes background estimation more 
uncertain, we chose PCU 2 as our ``standard'' detector to normalize the 
measured fluxes. 

\section{Data Reduction and Analysis}

We used the software package {\em HEASOFT 5.2},\footnote{see http://heasarc.gsfc.nasa.gov/docs/software/lheasoft} along with the associated
calibration files and background models, to reduce the {\em RXTE} data for
spectral analysis. Specifically, {\em Ftools 5.2} was used to extract 
spectra, both from PCA data and HEXTE data, and {\em XSPEC 11.2} was used
for spectral modeling. For timing analysis, however, we used custom 
software developed at MIT. The software is mature and has been used by the 
ASM team since the beginning of the mission. It was designed to work 
directly with raw packet data, instead of standard FITS files. 

\subsection{X-ray Spectral Analysis}
To minimize calibration uncertainties associated with the PCA response 
matrices, we limited ourselves to data from the first xenon layer of each 
PCU that was turned on during an observation. The trade-off is that we lost 
sensitivity to X-rays at energies roughly above 25 keV. We used the HEXTE
data to constrain the higher-energy portion of the X-ray spectra. It should 
be noted that cross flux-calibration between the PCA and the HEXTE remains 
an unresolved issue.\footnote{see http://lheawww.gsfc.nasa.gov/docs/xray/xte/crosscal.} The effect was taken into account in joint PCA/HEXTE 
spectral fits by making the relative normalization of the detectors an 
additional free parameter (i.e., multiplying the model by a constant in 
{\em XSPEC}). This approach has been commonly adopted (e.g., Coburn et al.
2002). Moreover, we chose to extract a
spectrum from each PCU and each HEXTE cluster separately and to perform a 
global fit for each observation, while floating the relative 
normalization of each spectrum (with respect to PCU 2). This allowed us to 
also take into account more subtle differences in the effective areas of 
individual detectors. It also added flexibility in our dealing with the
PCUs being turned on and off in the middle of an observation.

\subsection{X-ray Timing Analysis} 
To derive properties of the pulsed emission, we made a light curve for
each observation from the {\em Standard 1} data, which has a time resolution 
of 1/8 s and covers the entire PCA passing band. The 
light curve was then barycenter-corrected and folded at trial 
periods close to the expected value (since the pulse period is not expected 
to vary significantly from week to week). At each trial period ($P$), we 
compute the following statistic (Leahy et al. 1983):
\begin{equation}
S = \sum_{i=1}^{N} \frac{(r_i - \bar{r})^2}{\bar{r}} t_i,
\end{equation}
where $N$ is the number of phase bins, $r_i$ is the average count rate in 
phase bin~$i$, $t_i$ is the exposure time, and $\bar{r}$ is the overall 
average count rate. For this work, we chose $N=19$ and the folding epoch to 
be MJD 52466.0 (which is roughly at the middle of the monitoring campaign). 
In the absence of any periodic or secular variations, $S$ follows the 
$\chi^2$ distribution with $N$-$1$ degrees of freedom. 
 
The true pulse period should correspond to where $S$ peaks. To accurately 
locate the peak, we fitted the $S-P$ curve with a parabolic function in 
a small region around the peak. It should be noted that we did not subtract 
background from the light curves, in order to preserve simple statistics 
(Leahy et al. 1983). However, any non-Gaussian variation in the background 
rate would invalidate the basis for $S$ being $\chi^2$ distributed. 
Fortunately, the effects of such variation seem to be quite small in our
cases. A more serious complication arises from the fact that GX 1+4 itself 
exhibits aperiodic variations, as is evident by the presence of red 
noise in the power density spectrum (PDS; see Fig.~1 for an 
example). Therefore, $S$ is {\em not} a $\chi^2_{N-1}$ random variable here,
{\em even in the absence of a periodic signal}. 

To reveal the underlying distribution of $S$, we constructed $100$ 
simulated light curves for each observation, following a recipe in 
Timmer \& K\"{o}nig (1995). In order to derive analytic PDS models for the 
simulations, we excluded from a measured PDS a few bins around the harmonics 
of the pulsed signal (with the exact number of bins removed depending on
the exposure time of the observation), and subtracted off the Poisson 
noise (see, e.g., 
Zhang et al. 1996 for a description of the procedure). We then fitted the 
PDS continuum with a simple power law, which, by eye, seems to describe the 
data adequately. Note that in all cases except for one the short duration
of the observation does not allow the sufficient averaging of the PDS,
over multiple segments or frequency bins, to quantify the low-frequency 
noise with meaningful Gaussian errors. Consequently, we chose to seek best 
fits by following the least-squared algorithm, as opposed to the usual 
$\chi^2$ minimization. For the long observation ($\gtrsim 15$ ks), we 
binned the raw PDS by a factor of 10 and performed a $\chi^2$ fit. The 
best-fit model is quite similiar to that from the least-squared fit and 
the reduced $\chi^2$ of the fit is about 1.5.

The best-fit model was used to simulate light curves for each observation. 
Each simulated light curve was then re-randomized bin by bin (by replacing 
the number of counts, $n$, in a time bin with a random number drawn from a 
Poisson distribution with a mean value of $n$) to simulate noise due to
photon counting processes. The resulted simulated light curves were then 
folded at the measured pulse period to yield a value of $S$. Finally, we 
constructed a histogram of $S$ for that observation (in the presence of 
red noise and photon counting noise alone). Fig.~2 shows the 
68\% confidence regions of $S$ for all observations, along with the values 
measured from actual data. The results are fairly robust with respect to
modest amount of variation in the parameters of the input model.
Clearly, the presence of red noise can seriously affect our ability in 
estimating the significance of the pulsed signal.

\section{Results}

Fig.~3 shows the flux history of GX 1+4 in 2002. The source 
was quite bright at the beginning of the year and was seen to make a 
transition to a faint state in mid-June. The average flux is roughly an 
order of magnitude lower in the faint state
than in the bright state, although it varies significantly in both states. 
The minimum flux measured is about another order of magnitude lower, 
reaching $\sim 3\times 10^{-11} ergs\mbox{ }cm^{-2}\mbox{ }s^{-1}$. 

Toward the end of the year, GX 1+4 started to climb out of the faint state,
as its timing and spectral properties gradually revert to those of the 
bright state (see discussions below). Unfortunately, our monitoring 
of the source was terminated too soon to cover the entire transition, 
because of the proximity of the Sun to the source. 

\subsection{Pulse Period and Pulse Fraction}

X-ray pulsation is detected with high significance in all but four cases
in the faint state. Fig.~4 (top panel) shows the $S$--$P$ curve 
for one of the latter cases, which is basically flat. For comparison, we 
also
show in the figure results for an adjacent observation (taken about a week 
earlier), where the pulsation is clearly detected. The measured pulse 
periods (from the PCA data) are also shown in Fig.~3. Note 
that no period is shown for the
first observation, because it could not be reliably determined due to the
very short exposure time of the observation (less than 4 pulse cycles).
Although error bars are quite large in many cases (especially during the 
faint state), the spin-down trend can be seen from the figure. The best-fit 
spin-down rate is $4.4\times 10^{-8}\mbox{ }s/s$, which is less than half 
of the BATSE average (Bildsten et al. 1997). The source reached a pulse 
period of $\sim$138.7 s, the largest that has ever been measured. 

There are two sources that contribute to the uncertainty in measuring 
the pulse period: one is associated with the epoch-folding technique
and the other is due to the presence of red noise. The former is well
understood (see Larsson 1996), and we found that it was much smaller 
than the latter in all cases. We quantified the latter with simulations 
in the following manner. 
We added the observed pulsed signal to the simulated light curves (see 
\S~3.2), for each of the observations where the pulsation was detected
with high significance. We then derived pulse periods from the new 
simulated light curves through the same epoch-folding procedure. A 
histogram of pulse periods was constructed in the end and the 68\% 
confidence region was determined. Therefore,
the error bars have included the effects of intrinsic red noise of the
source as well as of statistical fluctuations. We caution, however, that
the reliability of the simulations depends sensitively on the accuracy 
of the input PDS model which we derived from the data. 

The observed pulse profile varies significantly from observation to 
observation (i.e., over a timescale of days), much more so in the faint
state. A thumbnail collection of pulse profiles 
are shown in Fig.~5. The profiles were derived by folding 
light curves (2--60 keV) with reference to the same epoch (MJD 52466.0), 
using the best-fit periods and period derivative (see Fig.~3). 
A feature that is common to a majority of the cases is the 
presence of a sharp dip in the pulse profile. The dips seem to occur at 
difference pulse phases. To be more certain about it, we experimented with 
deriving a phase coherence solution that connects the occurrence times of 
the dips, assuming they occur at a {\em constant} phase. In general, the 
pulse phase can be expressed as
$$ \phi (t) = \phi (t_0) + \nu_0 (t-t_0) + \frac{1}{2} \dot{\nu_0} (t-t_0)^2 + \frac{1}{6} \ddot{\nu_0} (t-t_0)^3 + \cdots $$
where $t_0$ is the reference epoch, and $\nu_0$, $\dot{\nu_0}$, and 
$\ddot{\nu_0}$ are the pulse frequency ($\equiv 1/P_0$), its first and 
second derivatives at $t=t_0$, respectively. We started by breaking the
longest observation (at MJD 52492) into six shorter segments. 
We folded the light curve of each segment and computed the time of the dip 
from the folded pulse profile. We did the same for two neighboring 
observations that were made about one week before and after the long
observation. We were able to derive a timing solution (up to $\ddot{\nu_0}$)
based on these data points. We then 
extended the time interval by adding one data point at a time and 
repeated the procedure. But, we failed to find a global phase coherent 
solution. Fig.~6 shows the residuals after the best-fit
model has been subtracted from the data. This result seems to suggest
that the dip does not always occur at a fixed pulse phase, although we
cannot rule out the possibility that it is caused by non-vanishing 
higher-order period derivatives. Similar dips were seen previously 
(e.g., Dotani et al. 1989; Giles et al. 2000), and were interpreted as 
being either due to obstruction of the ``hot spot'' by the accretion column 
(e.g., Giles et al. 2000; Galloway et al. 2001) or to resonance scattering 
of photons in the accretion column above the magnetic pole (Dotani et al. 
1989). However, it would seem difficult for either scenario to explain
why the dip does not occur at a constant pulse phase. 

As is obvious from Fig.~5, when GX 1+4 is relatively bright 
(the first 13 observations as well as the last few; see 
Fig.~3), its X-ray flux is strongly modulated.
When the source reaches the faint state, the pulsed fraction is significantly 
reduced and the pulse profile takes on drastically different forms. In 
four of the cases (see Fig.~5), the folded light curves 
show no sign of periodic modulation. It should be kept in mind that GX 1+4 
is known to exhibit red noise in its PDS (see Fig.~1 for an 
example). Consequently, much of the variation seen in the folded 
light curves can be caused by variations of the source that are associated 
with the red noise). To illustrate this point, we show, in Fig.~7, 
a comparison between a measured light curve and a simulated light curve 
for one of the observations in which no pulsation was detected. The 
non-Gaussian variation in the former can, therefore, be fully accounted 
for by the presence of red noise alone. We also investigated, with 
{\em goodXenon data}, the possibility that the signal might be detected 
only at higher energies for these cases. We constructed light curves that 
only include photons above roughly 7 keV and repeated the period-search 
procedure. No pulsation was detected. When the pulsation is not detected, 
the flux of GX 1+4 seems always to be 
$\lesssim 2 \times 10^{-10} ergs\mbox{ }cm^{-2}\mbox{ }s^{-1}$.
On the other hand, in a few other cases, the flux of the source falls 
below that threshold yet the pulsation is detected with high significance. 
Therefore, the phenomenology is not simple. 

To measure the amplitude of the pulsed signal, we defined the fractional
peak-to-peak pulsed fraction as $f_{pp} \equiv (F_{max}-F_{min})/F_{max}$,
and computed it for each observation. The results are shown in 
Fig.~8. Of course, the pulsed fraction thus defined is not 
expected to be zero, even in the absence of any pulsed signal, because of 
statistical fluctuations and aperiodic variabilities. The results from
simulations were again used to quantify the effects. Such ``baseline'' 
values of the pulsed fraction are shown in Fig.~8. Note that
the pulsed fraction does not stand above the baseline in some cases, 
even though the pulsation is clearly detected. This is due to the 
combination of uncertainties in the simulation, small pulsed fraction,
and poor statistics of the data.

\subsection{Spectral Characteristics}

The X-ray spectrum of GX 1+4 can be described by a model that consists 
of a cut-off power law and a Gaussian function. Also included in the model 
is the absorption along the line of sight. The observed continuum is, 
therefore, typical of accreting X-ray pulsars (White et al. 1983). The 
Gaussian component is needed to mimic the strong iron $K_{\alpha}$ line 
at $\sim$6.4 keV, which is known to be present in the X-ray spectrum 
of GX 1+4 (Kotani et al. 1999). This is the same model as the one that 
was adopted to study GX 1+4 during its 1996 transition to the faint state 
(Cui 1996) and thus facilitates direct comparisons of the results.

To account for residual calibration uncertainties, we followed the usual 
practice of including 1\% systematic error in the analyses. The fits are 
quite satisfactory, with the reduced $\chi^2$ values around $1$, except 
for a few cases at the beginning of the monitoring campaign when the 
source was bright. Even in the latter cases, the reduced $\chi^2$ values 
are never more than 
$\sim 1.7$, with deviations residing mostly at the lowest energies, where 
the PCA calibration is less certain.

When it is bright, GX 1+4 shows little intrinsic spectral variability. The 
measured photon
index falls in a narrow range of $1.1$--$1.3$; the exponential cut-off 
energy remains roughly at $8$ keV; and the e-folding varies in the range 
of $28$--$40$ keV. It is interesting to note that similar spectral 
characteristics were observed of the source during the last 3 
observations, indicating the emergence of GX 1+4 from the faint state. 

In contrast, the spectral properties of GX 1+4 appear to vary significantly 
in the faint state, on a timescale of days. Here, the photon index varies 
over a wide range 
of $0.2$--$2.0$, with no apparent correlation with the measured flux. 
The spectrum of the source is the steepest in cases where the 
pulsation is unmeasurable. In most cases, the spectrum rolls over at 
energies between $6$--$13$ keV, with the e-folding energy in the range 
of $11$--$45$ keV. However, no spectral roll-over is observed when the 
pulsation ceases. To illustrate the effect, 
we made a composite spectrum by combining data from the four observations 
in which no pulsation is detected, in order to achieve sufficient 
statistics at high energies. Fig.~9 shows the spectrum, along
with the best-fit model and the residuals. The presence of an iron 
$K_{\alpha}$ line is apparent; the continuum extends beyond 100 keV, 
following a simple power law with a photon index of roughly 1.6. To avoid 
divergence in overall luminosity, therefore, the spectrum must roll over 
at some energy, perhaps much beyond the HEXTE passing band. Very similar 
spectral variabilities 
were observed during the transition to a faint state in 1996 (Cui 1996). 

The infered hydrogen column density varies in the ranges of $4$--$9$ and 
$3$--$19\times 10^{22}\mbox{ }cm^{-2}$ for the bright state and faint state, 
respectively. It is not as well (sometimes, poorly) constrained in the 
latter case due to poorer statistics. These results are, again, comparable
to those obtained previously (Cui 1996). The variability might be related 
to the changing physical conditions in the wind of the M giant companion 
star.

\section{Discussion}

In this work, we have presented evidence for the cessation of X-ray 
pulsation of GX 1+4, when the source reached a low-flux 
state. While it is not technically feasible to rule out the presence of 
weak pulsation in such a case, we can, at least, conclude that the X-ray 
pulsed fraction must have decreased dramatically as the source made a 
transition from the bright state to the faint state. Similar phenomenon 
was observed previously (Cui 1996), but we have carried out a more 
detailed analysis in work, taking into account the presence of intrinsic 
red noise with Monte-Carlo simulations. We expect that the derived error 
bars on the pulsed fraction measurements, as well as significance 
estimates regarding the detection of pulsation, are more realistic. This 
is a point worth emphasizing, because much of the variability seen in a 
folded light curve could be attributed to the aperiodic secular variations 
of the source.

One thing that we do notice in this work is that there does not appear to 
be a fixed flux threshold below which the pulsation ceases, as one might
naively expect from the propeller effect. This complicates the 
interpretation. 
Perhaps, in the faint state, GX 1+4 was close to the transition threshold 
and it flipped and flopped stochastically between the propeller state and 
the non-propeller state. It might even be possible that there is some 
hysteresis associated with the transitions (see Fig.~2 of Lovelace, 
Romanova, \& Bisnovatyi-Kogan 1999). If the propeller phenomenon is 
indeed observed, it allows a direct estimation of the magnetic field in 
GX 1+4. 

\subsection{Propeller Effect}

For accreting pulsars, the determination of magnetospheric radius is still 
uncertain theoretically (e.g., Ghosh \& Lamb 1979; Wang 1996). The geometry
of the magnetic field is often assumed to consist of closed field lines, 
threading the accretion disk (Ghosh \& Lamb 1979). More recently, other 
field configurations (with open field lines originating in the accretion 
disk) have been investigated and may, in fact, be important for the 
propeller effect (e.g., Lovelace, Romanova, \& Bisnovatyi-Kogan 1999). 

At the Alfv\'{e}n radius, the magnetic pressure is balanced by the ram 
pressure of accreted matter. For a dipole field, the Alfv\'{e}n radius is
given by (Lamb, Pethick, \& Pine 1973)
\begin{equation}
r_A = 5.2\times 10^8\mbox{ }cm\mbox{ } L_{x,36}^{-2/7} B_{12}^{4/7} 
M_{1.4}^{1/7} R_6^{10/7},
\end{equation}
where $L_{x,36}$ is the bolometric X-ray luminosity in units of 
$10^{36}\mbox{ }ergs\mbox{ }s^{-1}$, $B_{12}$ is the field strength at 
the surface of the neutron star in units of $10^{12}\mbox{ }G$, and 
$M_{1.4}$ and $R_6$ are the mass and radius of the neutron star in units
of $1.4M_{\odot}$ and $10^6\mbox{ }cm$, respectively. The magnetospheric 
radius is probably a fraction of the Alfv\'{e}n radius, $r_m = \eta r_A$, 
where $0< \eta \leq 1$ (e.g., Ghosh \& Lamb 1979; Wang 1996; Lovelace, 
Romanova, \& Bisnovatyi-Kogan 1999).

The co-rotation radius, $r_{co}$, is defined as where the Keplerian velocity
is equal to the co-rotating velocity, $(GM/r_{co})^{1/2}=\Omega r_{co}$,
where $\Omega$ is the angular velocity of the neutron star. Therefore, we 
have
\begin{equation}
r_{co} = 4.5\times 10^9\mbox{ }cm\mbox{ }P_{138}^{2/3} M_{1.4}^{1/3},
\end{equation}
where $P_{138}$ is the rotation period of the neutron star, in units of
$138\mbox{ }s$. The propeller effect takes place when $r_{co}=r_m$. From 
equations (2) and (3), we find that the magnetic field strength is given 
by
\begin{equation}
B=4.4\times 10^{13}\mbox{ }G\mbox{ }\eta^{-7/4} P_{138}^{7/6} L_{x,36}^{1/2} 
M_{1.4}^{1/3} R_6^{-5/2}. 
\end{equation}
 
For GX~1+4, the observed 2-60 keV X-ray fluxes are 
$1.4$--$1.7 \times 10^{-10}\mbox{ }erg\mbox{ }cm^{-2}\mbox{ }s^{-1}$, 
for cases where the X-ray pulsation is undetectable. Assuming a distance of 
6 kpc (which is probably accurate to within a factor of 2; 
Chakrabarty \& Roche 1997), the observed luminosities are 
$\sim 7 \times 10^{35}\mbox{ }erg\mbox{ }s^{-1}$. Using this value in
Eq.~4, we already have $B > 3.7\times 10^{13}\mbox{ }G$. In some models, 
$\eta$ can be much less than 1 (e.g., $\sim 0.3$; Lovelace, Romanova, \& 
Bisnovatyi-Kogan 1999), so the inferred magnetic field might be an order of
magnitude larger. We must also make bolometric corrections, taking into 
account absorption at low energies and spectral roll-over at high energies,
the latter of which is necessary to keep the integrated flux from diverging.
While the low-energy correction is expected to be small ($\lesssim 30\%$),
the high-energy correction can be quite large, depending on where the 
spectrum rolls over, given the relatively flat spectrum of the source.
In extreme scenarios, therefore, the magnetic field of GX 1+4 could reach 
the magnetar regime!

\subsection{Origin of the Persistent Emission in the Propeller State}

In the propeller regime, the accreted matter can no longer be channeled to 
the surface of the neutron star by the magnetic field. Instead, it might 
be accumulated at the magnetosphere and gradually build up a dense envelope
around it. In this scenario, the matter might be accumulated preferentially
near the direction of the rotation axis and be expelled along the 
perpendicular direction (Wang \& Robertson 1985). In the process, the 
matter could be heated up to tens of millions of degrees and produce 
thermal bremsstrahlung radiation. However, the spectrum of such radiation 
should roll over exponentially at tens of keV, which we did not observe
in the case of GX 1+4. Therefore, the observed persistent emission is 
unlikely of magnetospheric origin. 

The propeller effect was also 
invoked to explain the large luminosity variation of 4U 0115+63 near its 
pariastron passage (Campana et al. 2001). It was argued that the emission
from the accretion disk would be important in the propeller regime. However,
in the case of GX 1+4, the inner edge of the disk would be quite far away 
from the neutron star in the propeller regime and thus be too cool to 
contribute much to the observed X-ray emission. Moreover, the observed 
X-ray spectrum of GX 1+4 is clearly not that of disk emission. 

On the other hand, we know that GX~1+4 has an M-giant companion star, which
is expected to produce a relatively dense, slow stellar wind (Chakrabarty
\& Roche 1997). As the propelled matter plows into this wind at a velocity
of the order of the escape velocity (Illarionov \& Sunyaev 1975), 
$(2GM/r_{co})^{1/2} \simeq 2000$ km/s, a strong shock might be formed. The
shock could then produce a population of relativistic electrons via the
first-order Fermi process. In this scenario, the observed 
persistent emission could simply be attributed to the synchrotron radiation 
of these electrons in the relatively strong magnetic field. The non-thermal 
nature of such emission is consistent with the lack of spectral roll-over 
in the propeller state. This emission mechanism is thought to be responsible 
for the unpulsed X--ray emission observed from the Be binary pulsar system 
PSR~B1259-63 near periastron (Grove et al. 1995).

One potential complication might arise from the fact that GX 1+4 is located
in the crowded Galactic center region. Since the flux of the source is quite 
low in the faint state, the issue of source confusion needs to be addressed. 
This can directly affect the derivation of pulsed fraction. More seriously, 
however, we need to be certain that GX 1+4 was detected at all in the 
faint state. We searched the existing X-ray catalogs for known sources 
within a $1^{\circ}$ radius circle around GX 1+4, and only found a number 
of very faint {\em ROSAT} sources, some of which might be transient in 
nature. For a Crab-like spectrum, the fluxes of these sources are all 
below $10^{-13}\mbox{ }ergs\mbox{ }cm^{-2}\mbox{ }s^{-1}$. This is 
consistent with results from our deep {\em Chandra} observation (with an 
exposure time of 60 ks) in the faint state: over 
the entire field of view ($8.3$\arcmin\ $\times$ 50.6\arcmin\ with the 
spectroscopic array), we detected only two very faint sources whose fluxes
are more than 2 orders of magnitude lower than the faint-state flux of 
GX 1+4. Moreover, the fact that we did also detect pulsation at comparable 
fluxes is re-assuring. The presence of the characteristic iron $K_{\alpha}$ 
line in all observations adds another piece of evidence that the source 
never completely turned off its X-ray emission during our campaign.

\acknowledgments
We wish to thank Duncan Galloway for many helpful discussions and John Finley 
for his advice on pulse timing and comments on the manuscript. This research 
has made use of data obtained through the High Energy Astrophysics Science 
Archive Research Center Online Service, provided by the NASA/Goddard Space 
Flight Center. We gratefully acknowledge financial support from NASA through 
an LTSA grant (NAG5-9998) and a {\em Chandra} grant (GO2-3052X).

\clearpage

\figcaption[f1.eps]{Sample power density spectrum of GX 1+4. The spectrum 
was made with data from the long observation taken on 2002 
August 6 05:50:00--14:30:00 (UT). The 
harmonics of the pulsed signal are clearly visible. Note the presence of 
strong red noise. } 

\figcaption[f2.eps]{Epoch-folding statistics. The shaded area represents 
68\% intervals 
of the statistic, derived from simulations (see text), when no pulsed signal 
is present. The actual measured values are shown in filled circles.} 

\figcaption[f3.eps]{Measured X-ray fluxes and pulse periods of GX 1+4 in 
2002. (top
panel) The fluxes were measured in the 2--60 keV band, with negligibly 
small error bars; the vertical bars at the top indicate observations in
which X-ray pulsation was not detected; and the dot-dashed line shows 
the average flux of the source in the faint state. (bottom panel) The 
error bars on the pulse periods were derived from simulations and represent 
68\% confidence intervals; and the dot-dashed line shows the best-fit 
spin-down trend. The boundaries of the faint state are crudely drawn in
dashed lines. For reference, MJD 52440 corresponds to 2002 
June 15. }

\figcaption[f4.eps]{Folding statistics at various trial periods for two 
observations about one week apart. The observations took place on 2002
July 15 and 21, respectively. No pulsation can be detected in the latter
(top panel), while it is clearly present in the other (bottom panel). }

\figcaption[f5.eps]{Observed pulse profiles of GX 1+4 in chronological 
order, from 
left to right and top to bottom. The average background rate has been
subtracted in each case. Note that no pulsation is apparent in 
four of the observations (as marked by asterisks).}

\figcaption[f6.eps]{Dip occurrence time residuals with the best-fit timing 
model (up to $\ddot{\nu}$) subtracted. Error bars are negligible on this 
scale. 
The dotted line indicates where the reference epoch ($t_0$) is (see text).
Note that not all observations clearly show the presence of a dip in the 
folded pulse profile (see Fig. 5).}

\figcaption[f7.eps]{Comparison between the measured (folded) light curve 
and a simulated light curve for the same observation. No pulsed signal was 
included in the PDS model for the simulations. The ``features'' in the
light curves are, therefore, purely due to the presence of red noise. }

\figcaption[f8.eps]{Fractional peak-to-peak pulsed fraction. The filled 
circles show the measured values. The shaded 
region represents 68\% intervals of the pulsed fraction, derived from 
simulations, when no pulsed signal is present. }

\figcaption[f9.eps]{Composite X-ray spectrum of GX 1+4 in the faint state. 
The spectrum was derived by combining data from those observations that
show no evidence of X-ray pulsation. The solid histogram shows the 
best-fit power-law model (which also includes interstellar absoption and 
an iron $K_{\alpha}$ line). Note that the spectrum extends beyond 100 keV, 
showing no evidence of rolling over. } 

\end{document}